\documentstyle[preprint,prl,aps,epsf]{revtex}
\tightenlines
\begin{document}
\draft
\title{KONDO RESONANCE EFFECT ON PERSISTENT CURRENTS THROUGH A QUANTUM DOT IN A 
MESOSCOPIC RING}

\author{V.Ferrari$^{a}$, G. Chiappe$^{a}$, E.V. Anda$^{b}$ and Maria A. 
Davidovich$^{b}$}

\address{$^{a}$Departamento de F\'\i sica, Universidad de Buenos Aires,\\
C.P.1428-Nu\~nez, Buenos Aires, Argentina\\
$^{b}$Departamento de F\'\i sica, Pontif\'\i cia Universidade
Cat\'olica do Rio de Janeiro\\
C.P. 38071, 22452-970 Rio de Janeiro RJ, Brasil} 
\date{\today}

\maketitle

\begin{abstract}
The persistent current through a quantum dot inserted in a mesoscopic
ring of length $L$ is studied. A cluster representing the dot and its
vicinity is exactly diagonalized and embedded into the rest of the ring.
The Kondo resonance provides a new channel for the current to flow. It is
shown that due to scaling properties, the persistent current at the Kondo
regime is enhanced relative to the current flowing either when the dot is
at  resonance or along a perfect ring of same length. In the Kondo regime
the  current scales as $L^{-1/2}$, unlike the $L^{-1}$ scaling of a
perfect ring. We discuss the possibility of detection of 
the Kondo effect by means of a persistent current measurement.     
 
\end{abstract}         
%\narrowtext

%\section{INTRODUCTION}

Electron transport through a quantum dot (QD) has been a subject of many 
experimental and theoretical studies in the last years. These small devices 
contain several millions of real atoms, but behave as if they were single 
artificial atoms.

Like real atoms they have a discrete spectrum of energy, which has been
measured \cite{yacoby1} and theoretically understood on the basis of a 
confinement potential and a full many-body electron-electron interaction 
treatment\cite{yo}.    

Unlike real atoms, electronic transport can be realized through a single QD. 
Experimental results\cite{Meirav} show  periodic oscillations of the conductance
 as a function of the electron density in the QD. These oscillations can be 
explained on the basis of a transport mechanism governed by Coulomb blockade
and single-electron tunneling\cite{Benaker}.
Thanks to the quantization of charge the effect of the Coulomb interaction
can  be understood in terms of a charging energy $\sim$ $\frac{e^{2}}{C}$
( where $e$ is the electronic charge and C the capacitance of the dot)  
necessary to add an extra electron to an already charged dot. This energy
is of the same order of the Coulomb repulsion $U$ between two electrons
inside the dot ${5}$.

Another manifestation of electron-electron interaction which was theoretically 
predicted \cite{Lee} some years ago is the Kondo effect in a QD coupled to
external leads. In this case however the effect is due to correlations between 
the electrons inside the QD and the conduction electrons in the leads. 
When the system operates in the so-called Kondo regime a  
resonance in the vicinity of the Fermi level, localized at the QD,
provides a  new channel for the mesoscopic  current to tunnel through,  
creating new phenomena which can be detected in a transport experiment. We have 
recently proposed \cite{anda} an experiment based on the Aharonov-Bohm
quantum interference phenomenon where the signature of the Kondo effect would
be clearly reflected on the current. In this work the system studied consists
of a ring connected to two leads having a QD inserted in one of its arms. 
Measurements of the current in a similar device have already demonstrated
[8]  the coherent character of the  electronic transport through the QD.
However, in this last experiment the size of the QD and the experimental  
temperature were not appropriate to observe the Kondo effect.
Its observation is a delicate task since it depends on several different energy 
scales and  their relative sizes, such as the coupling constant between the QD 
and leads $t'$, the Kondo temperature $T_{K}$ and the energy spacing
between  the dot levels $\Delta \epsilon$.
For example, if $t'^{2}/W>\Delta\epsilon$, where $W$ is the ring bandwidth,
the charge and energy quantization in the QD is lost and the Kondo effect  
disappears. On the other hand, diminishing $t'$ leads to an exponential
reduction of $T_{K}$ [9]. So, in order to get simultaneously accessible  
temperatures and charge-energy quantization $\Delta$$\epsilon$ must be large  
enough, which implies small size QD. Very recently, such a very small QD was  
obtained in a shallow two-dimensional electron gas heterostructure fabricated
by  electron-beam lithography. The Kondo effect was, for the first time,
observed  in this system, although as a small effect $[10]$.

Motivated by this experimental realization, we report in this letter the
study of persistent currents going through a QD embedded in a mesoscopic ring 
of length $L$ threaded by a magnetic flux. As we show below the persistent  
currents in these systems are enhanced by the Kondo effect, making its
detection possible. At very low temperature
the Kondo resonance $(KR)$  is a very sharp peak with width of the order of
$T_K$, commonly much sharper than the dot resonance of width $\Delta_d\sim
t'^{2}/W$. When the Kondo peak width is less than the energy spacing of the  
levels in the ring, that is $\Delta E > T_K$, it can be shown that the  
persistent current scales as $ L^{-1/2}$ unlike the $ L^{-1}$ scaling
which occurs in a  
perfect ring. This gives rise to an enhancement of the current intensity at
the Kondo regime. In this case the current peak could be larger than $Io$  $
(Io\sim ev_f/L$, where $v_f$ is the Fermi velocity), the intensity of the  
persistent current in a perfect ring $[11]$. This particular scaling of the  
current going through a very sharp resonance permits, in principle, a clear  
detection of the Kondo effect.

We study a two-level quantum dot embedded in a ring which is threaded by a  
magnetic flux and maintained at a fixed Fermi level. An external gate potential
$V_0$ is applied to the QD in order to be able to change its one particle
levels.  Part of this system, consisting of a cluster of $8$ atoms including
the  QD, is exactly solved by using a Lanczos algorithm. The cluster is then
connected to the rest of the ring. We calculate the persistent current in the  
ring, the charge inside the QD and the density of states projected on the QD. 

The system is represented by an Anderson-impurity  first-neighbor
tight-binding Hamiltonian. Although the electron-electron interaction exists
in the entire system we assume it to be restricted to the dot
where, due  to quantum confinement, the electrons interact more strongly. 
The total Hamiltonian can be written as
\begin {equation}
\label{eq1}
H=H_c+t \sum_{i,j\sigma} c^\dagger_{i\sigma} c_{j\sigma} + \hat T 
\end{equation}
where $H_c$ is the cluster Hamiltonian, $i$ and $j$ represent
nearest-neighbors atomic sites on the ring outside the cluster which interact  
through the hopping $t$, and $\hat T$ couples the cluster to the rest of the
ring. Denoting the two QD states by $\alpha$ and $\beta$ and numbering the
other six sites of the cluster from $\bar 1$ to $\bar 3$ and from $1$
to $3$, $H_c$ can be written as, 
\begin{eqnarray}
H_c&=&\sum_{\sigma\atop r=\alpha,\beta}(V_0+\varepsilon_r) n_{r\sigma}+U 
\sum_{r=\alpha,\beta} n_{r\sigma} n_{r\bar\sigma}+ U\sum_{\sigma\sigma'} 
n_{\alpha\sigma} n_{\beta\sigma'}\nonumber\\
&+&\left[t'\sum_\sigma(c^\dagger_{\alpha\sigma}+c^\dagger_{\beta\sigma}) 
(c_{1\sigma}+
c_{\bar 1\sigma} )+c.c.\right]+t \sum_{\sigma\atop m,n}
c^\dagger_{m\sigma}  c_{n\sigma}
\end{eqnarray}
and
\begin{eqnarray}
\hat T = t_{\phi} \left[c^\dagger_{\bar 3\sigma} c_{\bar 4\sigma} +  
c^\dagger_{3\sigma} c_{4\sigma}
\right] + c.c
\end{eqnarray}
\noindent
with 
\begin{eqnarray}
t_{\phi}=t\,e^{-i\pi(\phi/\phi_0)}
\end{eqnarray}
\noindent
where $U$ and $V_0$  correspond, respectively, to the electronic repulsion
and  the gate potential on the two states, $\alpha$ and $\beta$ with
energies  $\varepsilon_\alpha$ and  $\varepsilon_\beta$, respectively.  
The hopping matrix element between these two states and their nearest-neighbor
sites $1$ and $\bar 1$ is $t'$ and $m$ and $n$ stand for the other atomic  
sites of the cluster which interact through the nearest-neighbor hopping
$t$.  The external magnetic field producing the flux is incorporated in
the matrix elements between the cluster and the rest of the ring,
$t_{\phi}$ where $\phi$ is the magnetic flux crossing the ring,
$\phi=\oint \vec A\ \vec{dl}$, $\phi_0$ is the quantum of flux and  $\vec A$ 
is the vector potential of the external magnetic field.  

The ring is supposed to be in contact with an external reservoir of
electrons which fixes the Fermi level of the system. Therefore, a
variation of  the applied magnetic flux or the gate potential leads to a  
change of the total number of particles in the ring. 
To obtain the persistent current we calculate the Green function $\hat G$
for the complete system, within the chain approximation of a cumulant  
expansion\cite{Metzner} for the dressed propagators, by solving the Dyson
equation
\begin{equation}
\hat G = \hat g + \hat g\hat T \hat G.
\end{equation}
where $\hat g$ is the cluster Green function matrix obtained by the
Lanczos  method.
To take the charge fluctuation inside the cluster into account, we write  
$\hat g$ as  a combination of the Green function of $n$ and $n+1$
particles  with weights $(1-f)$ and $f$, respectively.
The Green function and the charge of the cluster can be written as,
\begin{eqnarray}
\hat g &=& \hat g_{n}(1-f) + \hat g_{n+1} f  \\
Q_{c}& =& (1-f)n + f(n+1) 
\end{eqnarray}
\noindent
The charge can also be expressed as,
\begin{equation}
Q_{c} =\int^{\epsilon_{F}}_{-\infty}\sum_{i} ImG_{ii}(w) dw
\end{equation}
\noindent
where $i$ runs over the cluster sites.
These equations are solved self-consistently in order to obtain $f$, $n$ and
 $\hat G$. The persistent current J and the charge in the dot $Q_d$ are  
obtained from,
\begin{eqnarray}
J = \frac{1}{2\pi}\int_{-\infty}^{\epsilon_{F}}Im\left[G_{i,i+1}(w)
-G_{i+1,i}(w)\right] dw 
\end{eqnarray}
\noindent
and
\begin{eqnarray}
Q_d =\int_{-\infty}^{\epsilon_{F}}Im\left[G_{\alpha,\alpha} (w) + 
G_{\beta,\beta} (w)\right] dw
\end{eqnarray}

Throughout this letter the energies are taken in units of $t$. To study
the  effect of the scaling on the persistent current we consider two rings  
with lengths $L = 200$ and $2000$, in units of the lattice parameter. 
According to our previous discussion we take a set of parameters
compatible  with an adequate Kondo temperature and a sharp charge
quantization.  The two dot level energies are chosen to be  
$\varepsilon_{\alpha}=0$  and $\varepsilon_{\beta}=2$ and $t'=0.05$. The
Fermi level is fixed at $\varepsilon_F=0$. The electron-electron
interaction and the magnetic flux responsible for the persistent current
are chosen to be $U=2$ and $\phi= \phi_0/4$. The physics of the problem is
not sensitive to the actual values of $U$ and $\phi$.

For $V_{0}> 0$ the levels  $\varepsilon_{\alpha}$ and
$\varepsilon_{\beta}$  are above the Fermi energy and the dot has no
electrons. The density of states of the system at the QD has two peaks at these
one-particle   
dot states. As $V_{0}$ is lowered  the  level $\varepsilon_{\alpha}$
goes through the Fermi level, permitting the entrance of electrons into
the QD, increasing the charge in the whole ring and driving the system into
the Kondo regime. This is shown in Fig.$1$ for $V_{0}=-0.3$ and $L = 2000$  
where the $KR$ appears as a sharp peak at the Fermi level and the dot
resonance as a wider peak below it. Notice also the presence of a splitted   
Coulomb peak separated from the dot resonance by an energy of the order of  
$U$.  In this situation a persistent current is installed in the ring. 

The current and the charge inside the dot are shown in Fig.$2$ as a
function  of the gate potential. The current has a peak at the dot
resonance ( $V_{0}\sim 0$ ) and, as $V_{0}$ decreases it changes its sign  
presenting a sharper and larger negative peak, as can be seen in Fig.
$2(a)$.  This last behavior reflects the entrance of the system into the
Kondo regime. The causes for the current sign change and the value of its
peak intensity will be discussed later. The current in this regime is due
to  the $KR$ which provides an extra channel for the electrons to go
through  the dot.  Although almost pinned to the Fermi level, the $KR$
moves slightly as $V_{0}$ decreases. This behavior, shown in the inset of  
Fig.$1$ for $L=2000$, is to be expected since the system does not possess  
particle-hole symmetry $[9]$. Due to the sharpness of the $KR$ its
movement through the Fermi level results in a very sharp peak in the
$I-V_0$ characteristic curve. Its intensity has a maximum at $V_{0}\sim -0.5$ 
when the $KR$ is at about $\varepsilon_F$. Decreasing further the gate
potential the current promoted by the Kondo effect persists up to the
point where the Coulomb peak aligns with the Fermi level and a second
charge enters into the dot, as shown in Fig. $2(b)$. This is reflected in
the current by the existence of another peak at $V_0 = -2.0$. The system
has a completely different behavior when $V_{0}$ is still further reduced.
The absence of a net spin when there are two electrons inside the dot  
inhibits the Kondo effect and the current goes to almost zero, as expected.

The current is an odd function with respect to $V_0 = -0.4$, as displayed
in the Fig.$2 (a)$, which reflects the electron-hole symmetry of the
system for this value of gate potential. The behavior of the dot charge as
a function of the gate potential explains the change of the current sign. 
When the Kondo resonance is partially above the Fermi level the charge
inside the dot is less than one, $(Q_{d} <1)$. The transit of the $KR$
through the Fermi level produces a small cusp in the curve of the dot
charge  at $V_{0}\sim -0.5$ and the current varies abruptly and changes
its sign. This is a consequence of the extra charge at the dot that,
according to Friedel sum rule \cite{friedel}, produces a phase  shift of
$\pi$ in the wave function at the Fermi level. This implies that the wave  
vector $k$ which contributes to the current changes sign during the
charging  process and so does the current.

The sharpness of the KR compared to the energy spacing of the ring, which  
restricts the participation of only one ring state in the current, permits
the  elaboration of an explanation  for the enhancement of    
the Kondo peak current relative to $I_0$ ( the current of a perfect ring
of the same size). The argument goes as follows: the persistent current of
a ring with a sharp Kondo localized peak weakly coupled to the
ring states can be calculated, within the one-particle framework provided by 
the mean field approximation of the slave bosons formalism \cite{read},
using degenerate perturbation theory. If the
localized level width is much smaller than the 
ring energy spacing, that is $\Delta_{d} << \Delta {E}$, this level is  
strongly hybridized only  with an almost degenerate ring state giving rise
to a bonding and anti-bonding states, which carry opposite currents. The
only situation in which a net current could go along the ring is when the  
bonding and the anti-bonding states are respectively below and above the
Fermi  level. The bonding state is therefore the only one participating in
the current.
In this limit it is straightforward to show that the
diagonalization of the $2\times2$ sub-matrix of the degenerated states at
the Fermi level gives a contribution to the current which scales as the
norm of the ring state, namely $L^{-1/2}$.
In a perfect ring, the $ L^{-1}$ dependence of the persistent current
results from the partial cancellation among 
the currents carried by the different states participating in the transport. 
On the other hand, if the ring localized level does not satisfy the
condition $\Delta_{d} << \Delta {E}$, perturbation theory is no longer
valid and the current has an intermediate scaling between the $L^{-1}$ and  
$L^{-1/2}$ regimes.
This behavior can be obtained by studying the persistent current of a ring  
with a dot at resonance as a function of the ring length $L$, for
different values of the dot resonant width $\Delta_{d}$ $\sim t'^2/W$.  
The results are displayed in Fig.$3$ where the persistent current maximum 
intensity as a function of the ring length $L$ is plotted in a Log-Log scale 
for several values of $t'$. The figure shows the $L^{-1}$ scaling for a  
perfect ring (t'=1), the $L^{-1/2}$ scaling for a weakly  connected dot  
(t'=0.01) and an intermediate behavior for other situations. Notice that
for a ring of length $ L=100 $ the current for $ t'=0.05 $ is about 0.4 of
the current of the perfect ring,  while for $L=2000$ the persistent
currents circulating along both rings have almost the same value.

The width of the $KR$ is of the order of the Kondo temperature $ T_{K}$ 
which decreases exponentially with the dot state energy relative to the
Fermi energy [9]. This allows the coexistence of very narrow peaks with  
moderate values of the  coupling between the ring and the dot. These  
conditions are consistent with having simultaneously a $ L^{-1/2} $
scaling and relatively high current intensities proportional to $t'$.  On
the contrary, in one-body problems the very small $t'$ required to obtain
a narrow peak reduces the current intensity.  
In a recent measurement of Kondo promoted mesoscopic currents [10], the
values of $T_K$  and of the width of the $KR$ were found to be of the
order of $0.01$ meV. If we consider a typical ring of length $L=2000$ the  
energy spacing of the ring  states satisfies $\Delta E >> T_K$ and so,
according to our above discussion, the persistent current in the Kondo
regime is enhanced relative to its value for a perfect ring and for a ring  
with a dot at resonance.
This enhancement can be clearly seen in Fig.$2(a)$. The maximum intensity
of the Kondo current ($V_0\sim -0.5$) is greater than $I_{0}$ and also
greater than the dot resonance current intensity ($V_0\sim 0$).  

The current intensity of a ring with $L=200$ is also shown in Fig.$2(a)$. As  
expected, it results to be larger than the one for $L=2000$. Although the  
mesoscopic nature of the persistent current appears for all gate
potentials, by increasing $L$ the Kondo current decreases much less than
the current circulating when the dot is at resonance. This is a
consequence of the already discussed different scaling behavior of these
two  regimes. 

Summarizing, we have analyzed the persistent current going through a
quantum dot embedded in a mesoscopic ring. Unlike the current of a perfect 
ring the persistent current in this system depends only upon the states at
the vicinity of the Fermi level. Our study shows that the Kondo resonance
provides a new channel for the electron to go along the system. As the Kondo 
resonance is very sharp the contribution to the current comes from the state 
at the Fermi level and the current scales as $ L^{-1/2} $.

The Kondo current results to be greater than the current of a perfect ring
of  same length. This property makes the detection of the Kondo effect in
this structure a feasible and very interesting possibility. 
 
We acknowledge Brazilian Agencies CNPq, CAPES and FAPERJ. One of us (G.
C.)  wishes to thank also to "Fundacion Antorchas" for finantial support.
(V. F) thanks CONICET for a Ph.D. fellowship.

\newpage 

FIGURE CAPTIONS:

\vspace{1cm}

FIG.1: Local density of states at the QD for $V_0=-0.3$ and a ring
length  L=2000. The inset shows the detail of the Kondo peak for three
values  of $V_0$

\vspace{1cm}

FIG2.: a) Persistent current as a function of $V_0$ through a QD
inserted in  a ring of size $L=2000$ (continuous line) and $L=200$ (dotted  
line) relative to the persistent current ($I_0$) of a perfect ring of size  
$L=2000$, for $t'=0.05$. (b) Charge at the QD as a function of $V_0$ for a  
ring with $L=2000$.

\vspace{1cm}

FIG. 3: Enhancement of the persistent current through a QD resonance
(without e-e correlation) relative to the persistent current of a perfect
ring of size $L=2000$ as a function of $L$, in a Log*Log scale, for four  
values of $t'$.

\vspace{1cm}

\begin{figure}
\epsfxsize=14cm \centerline{\epsfbox{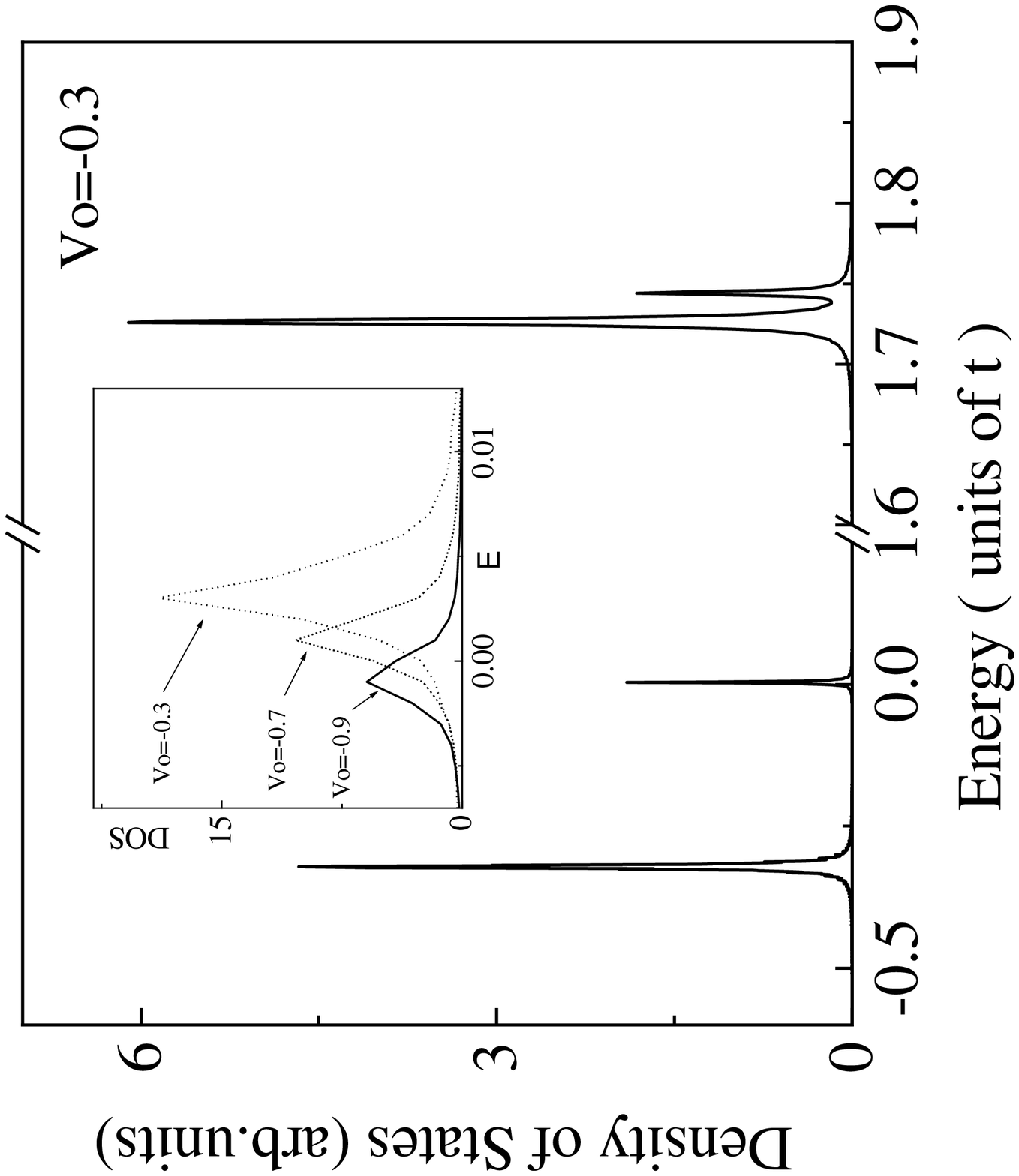}}
\end{figure}
\begin{figure}
\epsfxsize=14cm \centerline{\epsfbox{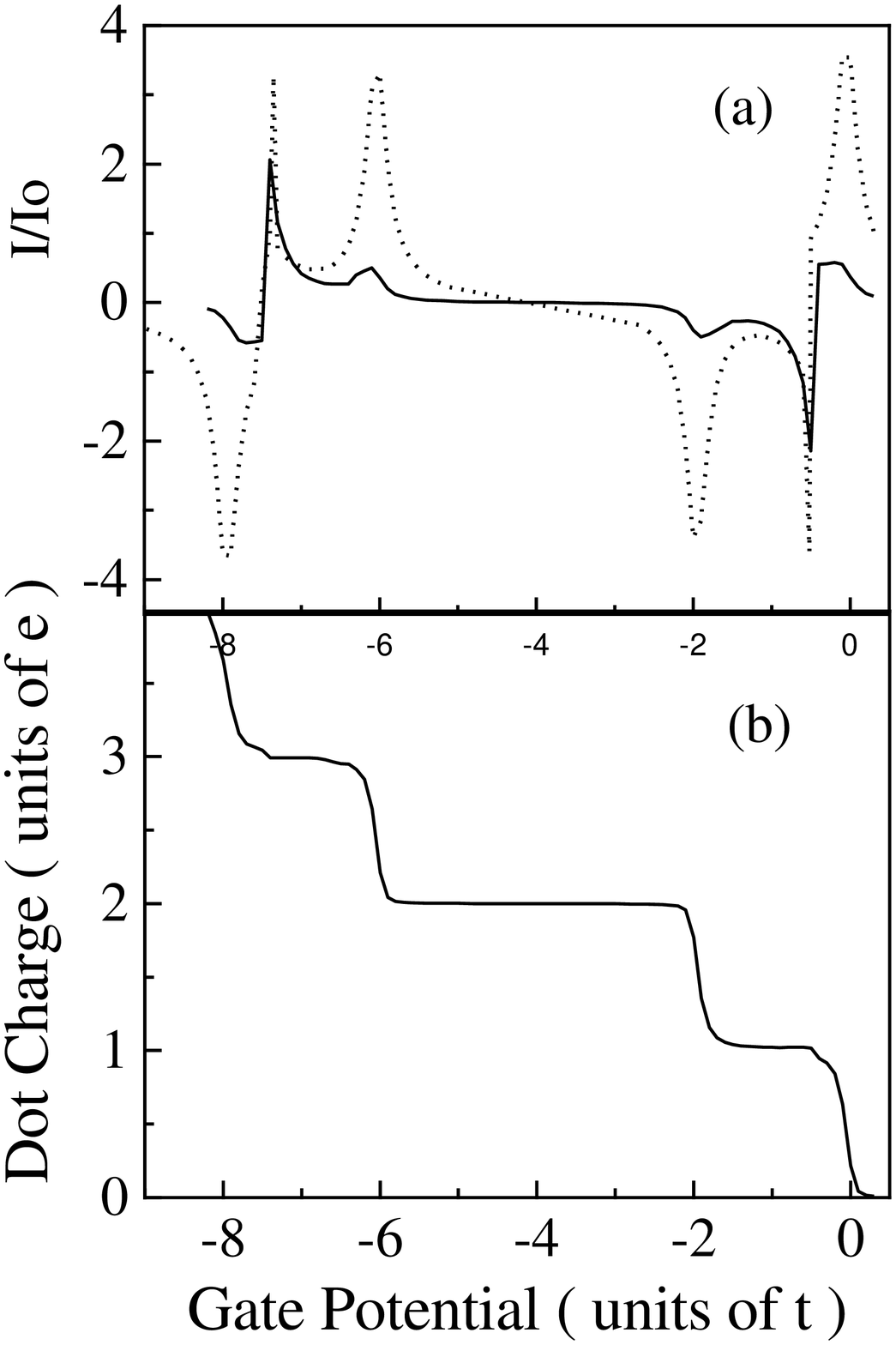}}
\end{figure}
\begin{figure}
\epsfxsize=14cm \centerline{\epsfbox{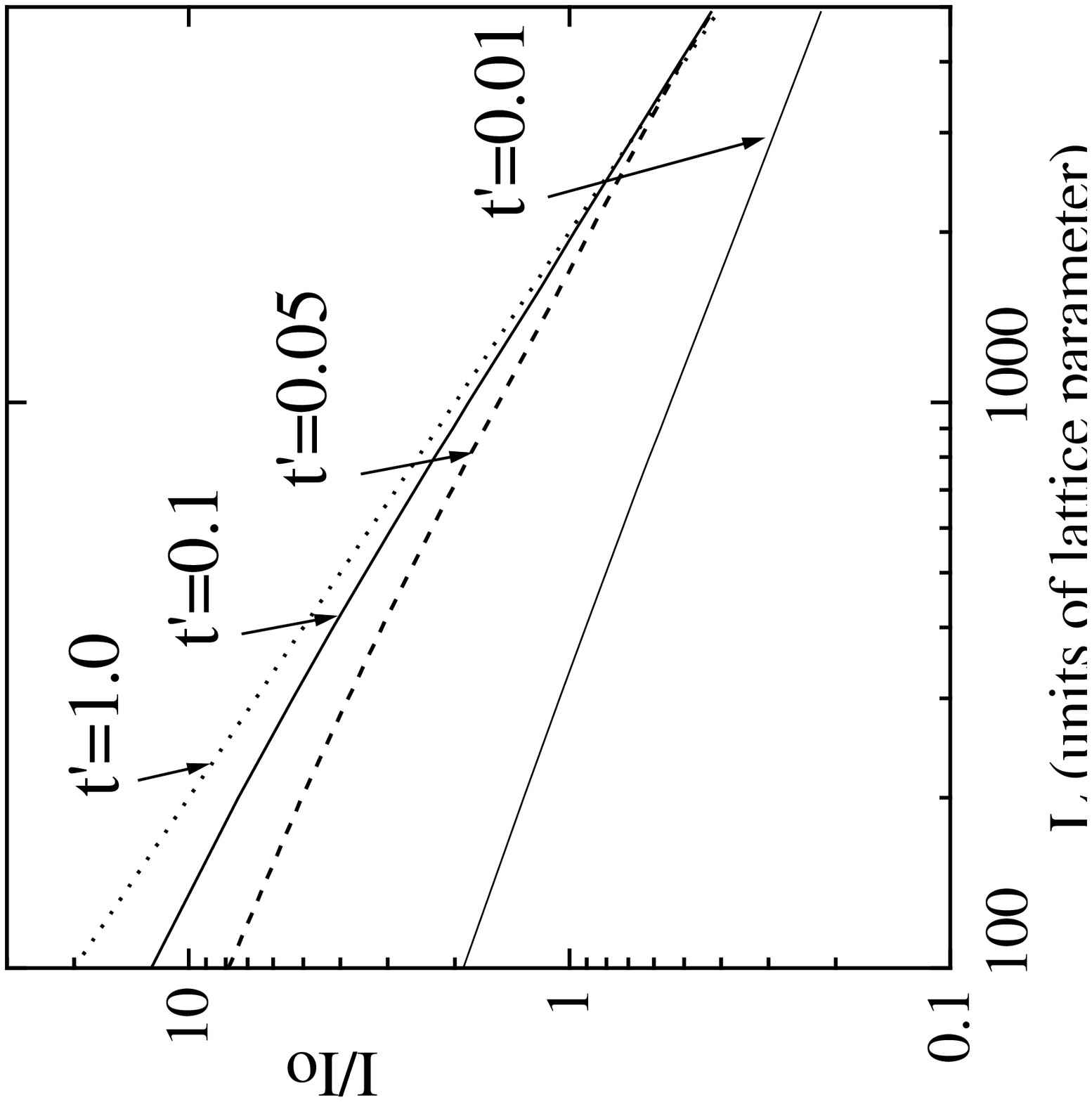}}

\end{figure}

\begin{references}
\bibitem{yacoby1} M. A. Kastner, Phys. Today $\bf 46$, 24 (1993); R. C.
Ashoori, Nature $\bf 379$, 413 (1996).\\    
\bibitem{yo} J. J.  Palacios, L. M. Moreno, G. Chiappe, E. Louis and C.
Tejedor, Phys.  Rev. B $\bf 50$, R5760 (1994).\\
\bibitem{Meirav} J. H. F. Scott-Thomas, S. B. Field, M. A. Kastner, H.I.
Smith and D. A. Antoniadis, Phys. Rev. Lett. $\bf 62$, 583 (1989).\\
U. Meirav, M. A. Kastner, M. Heiblum  and S. J. Wind, Phys. Rev. B $\bf
40$, 5871 (1989).\\
U. Meirav, M. A. Kastner and S. J. Wind, Phys. Rev. Lett. $\bf 65$, 771
(1990). \\
P.I. McEuen, E. B. Foxman, U. Meirav, M. A. Kastner, Y. Meir, N.S.
Wingreen and S. J. Wind, Phys.Rev. Lett. $\bf 66$, 1926 (1991).
\bibitem{Benaker} H. van Houten and C. W. J. Beenakker Phys. Rev. Lett
{\bf 63}, 1893 (1989).
\bibitem{U} A. Groshev, T. Ivanov and V.Valtchinov, Phys. Rev. Lett {\bf
66}, 1082 (1991). D.V. Averin, A. N. Korotkov and K.K. Likharev,  Phys.
Rev. B{\bf 44}, 6199 (1991). 
\bibitem{Lee} T.K. Ng and P. A. Lee, Phys. Rev. Lett {\bf 61}, 1768 (1988).
\bibitem{anda}  Maria A. Davidovich, E.V. Anda, J.R. Iglesias and G.
Chiappe, Phys. Rev B{\bf55}, R7335 (1997).
\bibitem{yacoby2}A. Yacobi, M. Heiblum, D. Mahalu and H. Shtrikman Phys.
Rev. Lett. {\bf 74}, 4047 (1995).
\bibitem{Hewson}  A. C. Hewson, "The Kondo Problem to Heavy Fermions",  
Cambridge University Press, 1993. 
\bibitem{nature}D. Goldhaber-Gordon, Hadas Shtrikmna, D. Mahalu,  David
Abusch-Magder and M.A. Kastner, Nature {\bf39}, 156 (1998)
\bibitem{Krieve} P. Sandstrom and I. V. Krieve, Phys. Rev. B {\bf 56},
9255  (1997). 
\bibitem{Metzner} W. Metzner, Phys. Rev. $\bf 43$, 8549 (1991)
\bibitem{friedel} J. Friedel; Phil. Mag. $\bf 43$, 152 (1952) \\
J. Friedel; Adv. Phys. $\bf 3$, 446 (1954)\\
J. S. Langer and V. Ambegaokar; Phys. Rev. $\bf 121$, 1090 (1961)
\bibitem{read} D. M. Newns and N. Read, Adv. in Phys. $\bf 36$,6, 799,
(1987)
\end{references}
\end{document}